\documentclass[12pt,preprint]{aastex}

\shorttitle{Proper Motions in SA~103 and an Orbit for the Virgo Stellar Stream}
\shortauthors{Casetti-Dinescu et al.}

\begin{document}

\title{Proper Motions in Kapteyn Selected Area 103: A Preliminary Orbit for the Virgo Stellar Stream}

\author{Dana I. Casetti-Dinescu\altaffilmark{1,2,3}, Terrence M. Girard\altaffilmark{1}, Steven R. Majewski\altaffilmark{4},
A. Katherina Vivas\altaffilmark{5}, 
Ronald Wilhelm\altaffilmark{6}, 
Jeffrey L. Carlin\altaffilmark{4}, 
Timothy C. Beers\altaffilmark{7}, and William F. van Altena\altaffilmark{1}}

\altaffiltext{1}{Astronomy Department, Yale University, P.O. Box 208101,
New Haven, CT 06520-8101, USA}
\altaffiltext{2}{Astronomy Department, Van Vleck Observatory, Wesleyan University, Middletown, CT 06459, USA}
\altaffiltext{3}{Astronomical Institute of the Romanian Academy, Str.
Cutitul de Argint 5, RO-75212, Bucharest 28, Romania}
\altaffiltext{4}{Department of Astronomy, University of Virginia,  P.O Box 400325, Charlottesville, VA 22904-4325}
\altaffiltext{5}{Centro de Investigaciones de Astronomia (CIDA), Apartado Postal 264, M\'{e}rida, 5101-A, Venezuela}
\altaffiltext{6}{Department of Physics, Texas Tech University, Lubbock, TX 79409, USA}
\altaffiltext{7}{Department of Physics and Astronomy, CSCE: Center for the Study of Cosmic Evolution, and JINA: Joint 
Institution for Nuclear Astrophysics, Michigan State University, E. Lansing, MI 48824, USA}


\begin{abstract}

We present absolute proper motions in Kapteyn Selected Area (SA) 103. This field
is located $7\arcdeg$ west of the center of the Virgo Stellar Stream (VSS,
Duffau et al. 2006), and has a well-defined main sequence representing the
stream. In SA 103, we identify one RR Lyrae star 
as a member of the VSS, according to its metallicity, radial velocity,
and distance. 
VSS candidate turnoff and subgiant stars have proper motions
consistent with that of the RR Lyrae star. The 3D velocity data imply an orbit
with a pericenter of $\sim 11$ kpc and an apocenter of $\sim 90$ kpc. 
Thus, the VSS comprises tidal debris
found near the pericenter of a highly destructive orbit. 
Examining the six
globular clusters at distances larger than 50 kpc from the Galactic center, and the
proposed orbit of the VSS, we find one tentative association, NGC~2419.
We speculate that NGC~2419 is possibly the nucleus of a disrupted system
of which the VSS is a part.
\end{abstract}


\section{Introduction}

It is now recognized that numerous stellar streams inhabit the
Galactic halo (e.g., Newberg et al. 2002; Belokurov et al. 2006; Grillmair et
al. 2006, based on the Sloan Digital Sky Survey - SDSS, and Vivas
\& Zinn 2006 based on the Quasar Equatorial Survey Survey Team - QUEST, to name only
a few). This newly discovered abundance of substructure has generated much
attention, as it qualitatively fits into the framework of a $\Lambda$ cold dark
matter cosmology that envisions the formation of the Galaxy via hierarchical
merging. However, the lack of available full phase-space descriptions of the substructure
limits our ability to properly model and quantify the merging history of our
Galaxy. It is certain that the formation and dynamical evolution of tidal
streams and overdensities is strongly affected by the Galactic potential, in
addition to the characteristics of the stream's progenitor (e.g., Johnston et al.
1999; Murali \& Dubinski 1999). These effects cannot be
properly modeled simply with projected positions on the sky, distances, and
in some cases radial velocities. Moreover, these tidal features cannot 
be properly disentangled in regions of the sky where they overlap in projection.
One such example is the region of the
Virgo stellar OverDensity (VOD). This region, first noted by
Newberg et al. (2002), and Vivas and Zinn (2003), covers a large region
on the sky ($\sim 1000\arcdeg$), as recently shown by Juric et al. (2008). While
a rather low surface-brightness feature ($ < 32.5$ mag~arecec$^{-2}$), the VOD
itself appears to have substructure seen in surface density as well as in
radial-velocity space (Duffau et al. 2006 - D06, Newberg et al. 2007 - N07; Vivas et al.
2008 - V08; Keller et al. 2008 - K08;  Keller et al. 2009; Prior et al. 2009 - P09). The
best-characterized stream in the VOD is known as the Virgo Stellar Stream
(VSS, a.k.a. the 12.4-hour clump of Vivas \& Zinn 2003) and it is the subject of this paper.
The VOD remains a complex region, where stars from both the leading and trailing tidal tails of Sgr
may be found as noted by Kundu et al. (2002), Mart\'{i}nez-Delgado et al. (2007), P09.

In this paper we present absolute proper motions in Kapteyn Selected Area (SA)
103, located at (R.A., Dec.) = $(178.8\arcdeg, -0.6\arcdeg)$ and (l, b)$ =
(264.6\arcdeg, 59.2\arcdeg)$. The center of the VSS is at (R.A., Dec.)$ = (186\arcdeg,
-1\arcdeg)$; it extends in R.A. from $175\arcdeg$ to $200\arcdeg$ (D06), 
while in Dec. it has been shown to extend from $0\arcdeg$ to $-15\arcdeg$
(P09). SA~103 is thus $\sim 7\arcdeg$ west of the VSS center as defined by D06.
In this field, we were able to
identify one RR Lyrae star as a VSS member for which the absolute
proper motion is measured. This allows us to obtain a preliminary orbit for the
VSS.

\section{Data and Results}

SA~103 is one field of $\sim
50$ from the proper-motion survey described by Casetti-Dinescu et al. (2006,
CD06). In CD06 all details concerning the reduction process and the
derivation of proper motions are presented. Here, we only briefly mention the
basics of this survey. Each field covers $40\arcmin\times40\arcmin$,
and makes use of photographic plates taken at three different epochs. The modern
epoch consists of plates taken between 1996 and 1998 with the Las Campanas du
Pont 2.5m telescope, the intermediate epoch consists of Palomar Observatory Sky
Survey plates (POSS-I) taken in the early fifties with the Oschin
Schmidt 1.2m telescope, and the old epoch consists of plates taken between 1909
and 1911 with the Mount Wilson 1.5m telescope. The modern and old
plates were measured with the Yale PDS microdensitometer. For the POSS-I plates
we have used scans done by both the Space Telescope Institute (the Digitized Sky
Survey) and the US Naval Observatory. For SA~103 we have used two du Pont plates
two overlapping POSS-I fields, and one 60-inch Mt. Wilson plate. The region of
SA~103 has complete SDSS DR7 (Abazajian et al. 2009) coverage. Thus, by
comparing with the DR7 data, our proper-motion catalog in SA~103 is $86\%$
complete at $g=20.0$, and becomes $15\%$ complete at $g = 21.0$. The limiting
magnitude for objects that have over a 80-year baseline (i.e., those that were
measured on the Mt. Wilson plates) is $g \sim 20.0$, with $\sim 90\%$
completeness at $g = 17.0$. For objects that are well-measured (i.e., $g \le
18$), we obtain proper-motion errors of $\sim 1$ mas~yr$^{-1}$, as shown in
CD06 and Casetti-Dinescu et al. (2008). All proper-motion units in this paper are mas~yr$^{-1}$.

We have searched the Vivas et al. catalogs to
find matches with stars in our SA~103 field. We have identified one RR
Lyrae star -- RR~167, classified as Bailey type ab, in the Vivas et al. catalogs in our field.
RR~167 ((R.A., Dec.) = $(178.893\arcdeg, -0.600\arcdeg)$) 
is located near the center of our field and, with an average magnitude $V = 16.77$, it is a
well-measured star, with the proper motion derived from thirteen position
measurements across 86 years. The same star is also listed in the K08
catalog. Vivas et al. (2006) determine a heliocentric distance
of 16.9 kpc, while K08 a distance of 18.3
kpc. Both studies estimate a $7\%$ uncertainty in the distances, therefore the
distances agree within uncertainties. No radial velocity (RV) or
metallicity information is provided for RR~167 in subsequent studies that
focus on the VSS. We have thus searched the SDSS DR7 data
for spectroscopic observations in SA~103, and fortunately we have found that RR~167
had two spectra taken near minimum light (phase 0.70 and 0.75). 
The phases of the spectroscopic observations were calculated using the QUEST ephemeris which,
for this star, was based
on 30 epochs in the light curve. From these spectra, the systemic
heliocentric radial velocity of the star was determined to be $242.7 \pm
14.0 $ km~s$^{-1}$ using the procedure described in V08,
which consists of fitting the radial velocity curve of the well studied RR
Lyrae star X Arietis. The 2 SDSS radial velocities for RR~167 fit quite well
the curve of X Arietis (rms of the fit = 3  km~s$^{-1}$). The quoted
error however includes the uncertainties due to possible variations from star to
star in the slope of the radial velocity curve and the phase when the
systemic velocity occurs. In the Galactic rest frame the
velocity is $V_{gsr} = 134.3\pm 14.0$ km~s$^{-1}$. 
The metallicity was determined from the CaII K and H$\delta$, H$\gamma$, and
H$\beta$ equivalent width using the procedure outlined in Layden (1994).
The metallicity values derived from the two spectra are
$-1.79\pm0.22$ and $-1.84\pm0.21$. 
Based on this metalicity ($-1.8\pm0.2$) we have re-estimated the distance to RR~167,
and we obtain $17.0\pm0.9$ kpc, which is the value that we use for the orbit determination.
The properties of stars in the VSS as determined by previous studies are summarized below. 
D06 find a somewhat smaller mean velocity than
 subsequent studies, $V_{gsr} \sim 99$
km~s$^{-1}$ from a sample of 6 RR Lyrae and blue horizontal branch (BHB) 
stars\footnote{However, a recent revision of the QUEST velocities gives a
mean velocity for the VSS of 126 km~s$^{-1}$, in perfect agreement with the
other works (Duffau 2008)}. The metallicity determined by D06
is [Fe/H] = $-1.86\pm0.15$, with the VSS having a significant metallicity spread
($\sim 0.4$ dex) intrinsic to the structure. 
N07 obtain $V_{gsr} = 130\pm10$ km~s$^{-1}$ from data on F-type turnoff star, 
while P09 obtain
$V_{gsr} = 127\pm10$ km~s$^{-1}$ (average of four RR Lyrae stars), and [Fe/H] = $-1.72$
and $-2.15$ for two stars with available metallicity estimates. The distance to
stars in the VSS sampled by D06 is 19 kpc, while N07 estimate a distance of 18 kpc. 
From the more recent study by
V08, the distance to VSS ranges between 12 and 19 kpc. 
Thus, the estimated distance, RV, and metallicity of RR~167 make it a very likely
member of the VSS.

Figure 1 shows the formal proper-motion errors as a function of $g$ magnitude
for stars in SA~103. 
RR~167 (filled circle) has formal errors within
the range for stars at that particular magnitude. The proper-motion zero-point
for an inertial reference frame is determined from the measurement of background
galaxies as classified by SDSS. We use 302 galaxies with measured proper motions less
than 20 mas~yr$^{-1}$ and that had at least four positional measurements. To
determine their mean and dispersion, we apply the probability-plot method
(Hamaker 1978) using the inner $80\%$ of the proper-motion distribution. In our
data we also have four rather faint QSOs (as identified from SDSS DR5, see
Schneider et al. 2007), $g = 18.8 - 19.5$. The galaxies and QSOs determinations
agree within errors. We obtain a final zero point from the mean proper motions
of the galaxies and the QSOs, weighted by their formal errors. Our final zero
point is $\mu_{\alpha}cos\delta = 3.69 \pm 0.27$ mas~yr$^{-1}$ and $\mu_{\delta}
= 2.62 \pm 0.28$ mas~yr$^{-1}$. Figure 2 shows absolute proper-motion diagrams
(i.e., with the zero point applied such that galaxies are centered on 0,0) for
stars (left panel) and galaxies (right panel). RR~167 is the filled circle. The
open circles in the galaxies' diagram are the QSOs. The absolute proper motion
of RR~167 is $\mu_{\alpha}cos\delta = -4.85 \pm 0.85$ mas~yr$^{-1}$ and
$\mu_{\delta} = 0.28 \pm 0.85$ mas~yr$^{-1}$. Here, the error is determined by
adopting the median value for the error of a star with the magnitude of RR~167, as seen
in Fig. 1 (i.e., 0.8 mas~yr$^{-1}$), and adding in quadrature the error in the
zero point determination. This is done to avoid underestimating the error by
using the value formally obtained from the fit of positions as a function of
time. Clearly, RR~167 has a high proper motion, detected at the $5.7\sigma$
level. Such a large proper motion for an object at $\sim 17$ kpc implies a very
energetic orbit (see next Section). The proper motion corrected for the reflex solar motion is $\mu_{\alpha}cos\delta = -3.50 \pm 0.85$ mas~yr$^{-1}$ and
$\mu_{\delta} = 2.33 \pm 0.85$ mas~yr$^{-1}$.

To further explore the presence of VSS debris in SA~103 we present observed
color-magnitude diagrams (CMDs) and proper-motion diagrams, and compare with
their Besancon-model (Robin et al. 2003) equivalents. The top panels of Figure 3
show the CMDs of an area centered on SA~103, and covering our
$40\arcmin\times40\arcmin$ area. The left panel shows all stars from DR7, the
middle panel shows the Besancon model data, and the right the DR7 data for stars
with available proper-motion measurements. The location of RR~167 is highlighted
with a filled circle. The dashed line indicates the approximate limiting
magnitude of our proper-motion data.
The DR7 data exhibit a distinct main-sequence like overdensity, as well as a
mild enhancement at the location of a corresponding horizontal branch. The
SA~103 region has also been analyzed by P09, who examined SDSS
data as well as their own $V,R$ starcounts, and found a clear stellar excess
compared to the Besancon-model starcounts in the turn-off region of the CMD
(their Figs. 13 and 14 - top panels). They also compared the region in SA~103 to
other two regions, of which one, at R.A. $= 218\arcdeg$, shows no stellar excess
compared to the Besancon model. 
To better characterize the stellar overdensity in SA~103, we overlay the
fiducial sequence of globular cluster M~53 in Fig. 3. The fiducial
sequence is constructed from the DR7 data, and is represented with a grey line
in the top-left panel. M~53 has a metallicity [Fe/H] $= -1.99$, and is located
at a distance of 17.8 kpc (Harris 1996). Thus, it is representative of the VSS
population. The reddening in SA~103 is E(B-V) = 0.026, while at the location of
M~53, it is E(B-V) = 0.021 (Schlegel et al. 1998). Therefore, only small adjustments
due to reddening
had to be made to the fiducial sequence of M~53 in order to align it with the
VSS. We chose not to adjust the distance modulus, since our distance to RR~167
agrees with that of M~53 within errors. 
Figure 3 shows that the VSS closely follows M~53's fiducial
sequence; it is slightly redder, and there are no BHB stars.
This indicates that the VSS is slightly more metal rich than M~53, in agreement
with the spectroscopic determinations that give a 0.2dex difference. 

The next step is to identify more likely VSS candidates in our proper-motion
field. In the well-measured regime ($\sim 1$ mas~yr$^{-1}$, $g < 18.0$) we expect
very few stars. Both K08 and Vivas et al. (2004) have surveyed this region, and
found only one RR Lyrae star which is in keeping with our expectations 
from the proper-motion field.
There could be several more red
horizontal-branch stars, as the CMD appears to indicate, however without RV
information it is difficult to confidently distinguish them from field stars.
Likewise, we expect very few giants. From M~53's DR7 CMD (within a 10-arcmin
radius from the cluster center) we obtain a ratio of about three red giant-branch stars
(approximately from the tip to the subgiant region) for each horizontal-branch
star. Thus, even along the giant branch we can't expect more than about ten VSS
stars. Lacking RV information, the giant branch VSS stars have the additional problem of being
virtually impossible to distinguish from the $\sim 60$ field stars that inhabit the same band
as the M~53 giant branch.
The highest contrast between the VSS and the
Galactic field where proper-motions are available is the subgiant and turnoff
region, which is also where proper-motion errors are the largest. Nevertheless,
we have selected stars in this region to check the overall proper-motion
distribution, and compare it to that predicted by the Besancon model. The stars
are selected within the rectangular box highlighted in the CMDs of Fig. 3. Our
data include 93 stars, while the Besancon model predicts $\sim 66$ stars. 
Note that while the
Besancon sample is complete, the data are not (at $g = 20$, completeness is
$85\%$). The proper-motion distribution of our data is shown in the bottom-left
panel of Fig. 3, where formal error bars are included for each star. 
The middle panel shows the proper-motion density contour map of our data, while the left
panel shows that of the Besancon data convolved with 4 mas~yr$^{-1}$ error. 
Reassuringly, in the observed data the region of peak density includes RR~167.
The Besancon proper-motion distribution shows a density peak at a different location than that of 
RR~167; this peak is also lower than that of the observed distribution, which 
reflects the stellar overdensity found in this area.

\section{A Preliminary Orbit Determination for the VSS}

The velocity components in cylindrical coordinates $(\Pi, \Theta, W)$ are
calculated using the RV, distance and proper motion listed above for RR~167.
Using the Dehnen et al. (1998) peculiar velocity of the Sun, at R$_0$ = 8 kpc, and $\Theta_0 = 220$
km~s$^{-1}$, we obtain $(\Pi, \Theta, W) = (253\pm65, 221\pm67,142\pm37)$
km~s$^{-1}$. We have integrated the orbit in the Johnston et al. (1995) Galactic
potential, following the procedure described in Dinescu et al. (1999) to
estimate uncertainties in the orbital parameters. We obtain an orbit with a
pericentric radius of $11\pm1$ kpc, an apocentric radius of $89_{-32}^{+52}$
kpc, and an eccentricity of $0.78\pm 0.06$. The orbital inclination is
$58\arcdeg\pm5\arcdeg$, derived as $90\arcdeg$ - sin$^{-1}(L_z / L)$, where
$L_z$ is the angular momentum along the direction perpendicular to the Galactic
disk, and $L$ is the total angular momentum. The radial orbital period is
$1.2_{-0.4}^{+0.6}$ Gyr. These orbital elements rule out any association of the
VSS with Sgr's orbit (e.g., Dinescu et al. 2000; Dinescu et al. 2005). In
particular, examination of the three integrals of motion for the orbits of Sgr
and the VSS shows that while the total orbital energy and angular momentum are
not determined with sufficient precision for the VSS to safely rule out its
association with Sgr, the $L_{z}$ differs at the $2.2\sigma$ level. 
The current location of the RR~167 is $\sim 17$ kpc from the Galactic
center. Thus, the orbit obtained indicates that the stars in the VSS are near
pericenter. Interestingly, the smallest distance for stars in the VSS is
estimated to be 12 kpc (V08). Using this distance at the sky 
location of RR~167, we obtain a Galactocentric distance of 14 kpc, which is in 
reasonable agreement
with our determination for the pericentric radius of the orbit. Although the
uncertainty in the apocentric radius is rather large, owing to distance and
proper-motion errors, it is apparent that the orbit is very eccentric and thus
rather destructive for the parent satellite. Disruption models of satellites on
eccentric orbits indicate that tidal debris piles up preferentially at the
turning points of an orbit (e.g., the
disruption of Sgr: Law et al. 2005).
Therefore, the VSS structure is consistent with a pericentric density enhancement
of a disrupted satellite on a highly eccentric orbit. 
While the orbit is somewhat uncertain, it is a highly energetic one, with an
apocenter beyond 50 kpc. We show this orbit in Figure 4, along with that of Sgr for comparison.
Since there are only six globular clusters beyond a
Galactocentric radius of 50 kpc, we searched for positional coincidence between
the orbit of the VSS and distant clusters.
There is only one cluster that can be associated with this orbit, namely
NGC 2419 which is 13 kpc away from the closest point in the VSS orbit, integrated 
back in time for 5 Gyrs.
NGC~2419 is a well-known massive cluster 
with a low metallicity ([Fe/H] $= -2.12$), and a very
extended BHB. As such, its properties fit the idea that
massive, extended BHB clusters are the nuclei of disrupted dwarf
galaxies (Lee et al. 2007). 
Cluster NGC~2419 ((l,b) = $(180.4\arcdeg, 25.2\arcdeg) $
has been tentatively associated with debris from Sgr by Newberg et al. (2003).
This association is based on the proximity of NGC~2419 to Sgr's orbital plane, and on the
finding of an overdensity of A-type stars (assumed to be BHB stars) 
at a galactocentric distance of $\sim 90$ kpc
that lie near Sgr's orbital plane toward the direction of NGC~2419 (Newberg et al. 2003).
Disruption models of Sgr (Law et al. 2005) show that debris from Sgr can be found 
in the region and at the galactocentric distance of NGC~2419, as part of the trailing tail,
thus suggesting that the cluster and the A-type star overdensity could belong to Sgr.
The low radial velocity of NGC~2419, $V_{gsr} = -27$ km~s$^{-1}$ (Harris 1996)
indicates that the cluster is at a turning point in its orbit (in this case the apocenter), 
which can be equally implied by the membership to Sgr and VSS (Fig. 4).
Since NGC~2419 is in the Anticenter direction, 
only its transverse velocity can establish whether it is associated to 
any of these two streams.

\acknowledgments

Financial support from NSF grants AST-0406884 and AST-0407207 for this research
is acknowledged. T.C.B. acknowledges partial support from PHY 08-22648: Physics
Frontier Center/Joint Institute for Nuclear Astrophysics (JINA), awarded by NSF.
This publication makes use of SDSS data
products. Funding for the SDSS and SDSS-II has been provided by the Alfred P.
Sloan Foundation, the Participating Institutions, the NSF, the US Department of
Energy, NASA, the Japanese Monbukagakusho, the Max Planck Society and the Higher
Education Funding Council for England. The SDSS Web site is
http://www.sdss.org/.

{\it Facilities:} \facility{SDSS}.

\clearpage

\begin{figure}
\includegraphics[scale=.70]{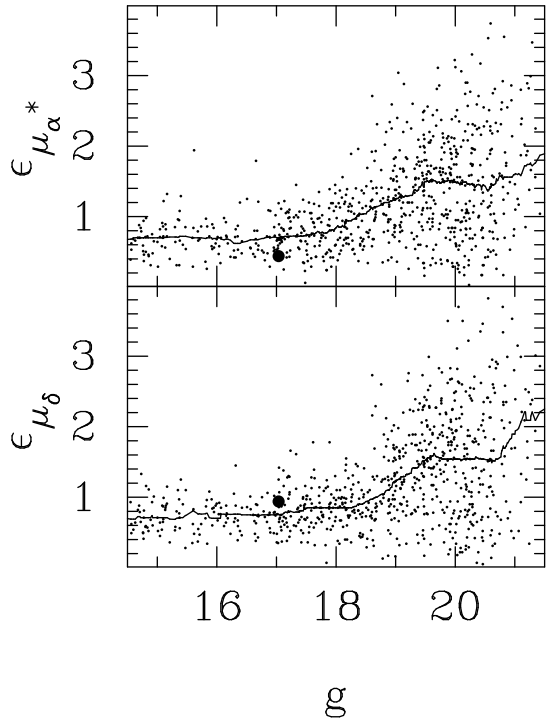}
\caption{Proper-motion errors in each coordinate 
as a function of magnitude for stars. The line represents the moving median
of the sample. The filled circle is RR~167. Here, $\mu_{\alpha}^{*} = \mu_{\alpha}~cos~\delta$.
\label{fig1}}
\end{figure}

\clearpage

\begin{figure}
\includegraphics[scale=.70]{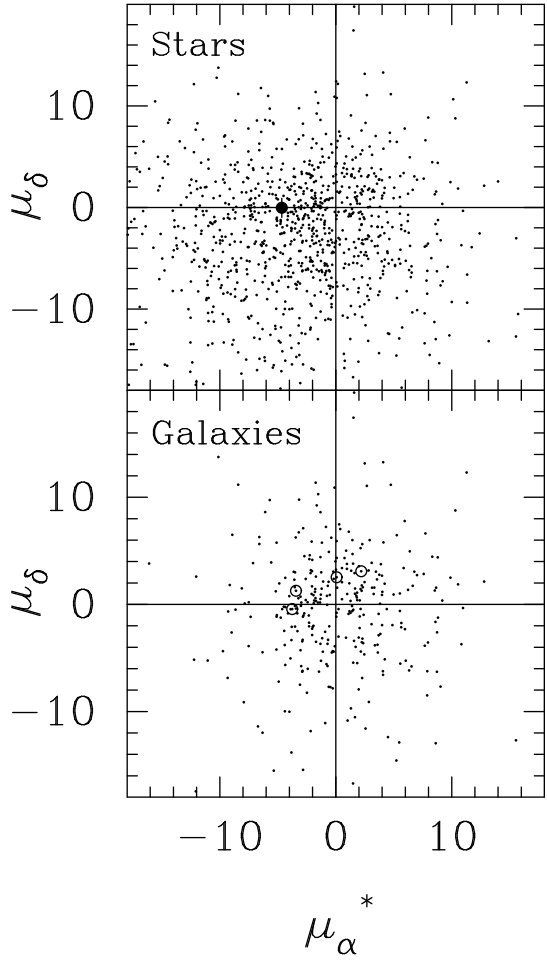}
\caption{Absolute proper-motion diagrams for stars (left) and galaxies (right).
The filled circle is RR~167. The open circles in the right panel show
the four quasars that were measured in this area. \label{fig2}}
\end{figure}

\begin{figure}
\includegraphics[angle=-90,scale=.60]{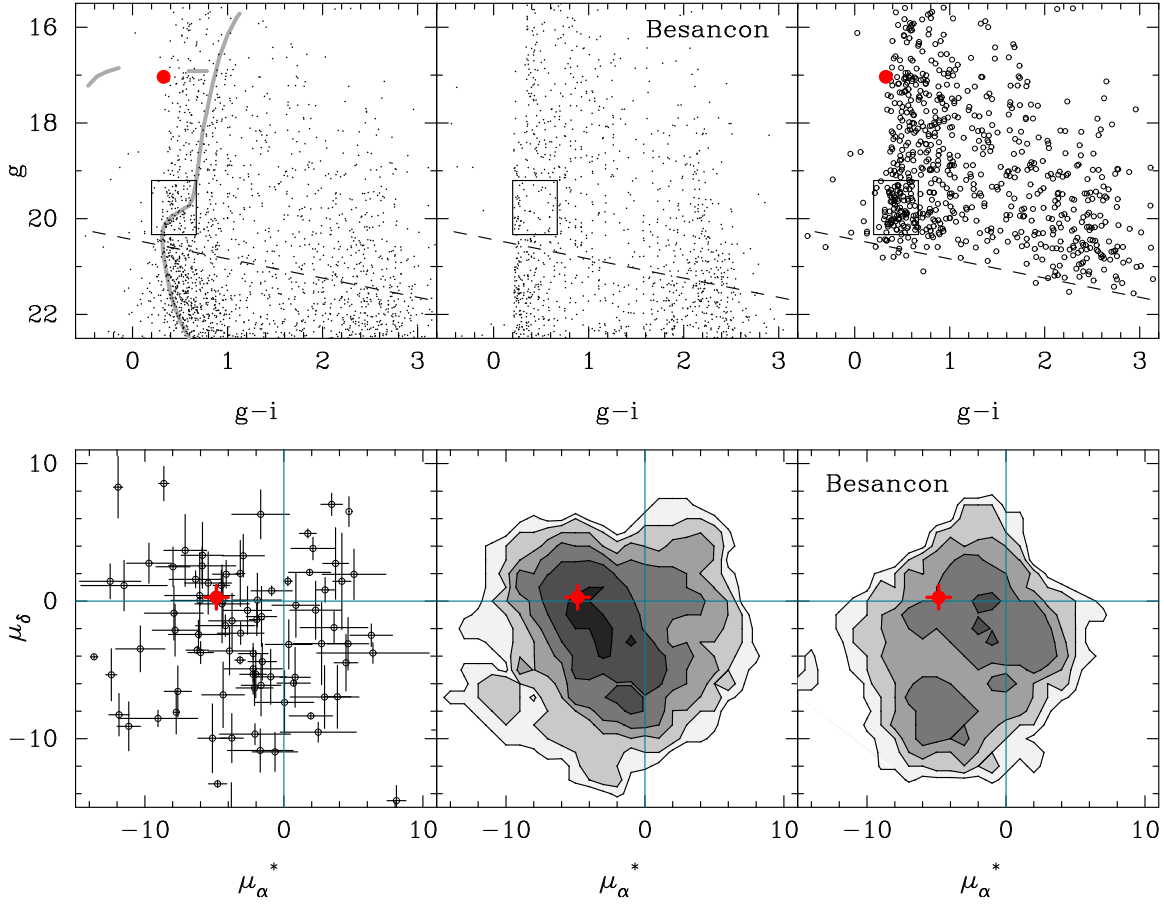}
\caption{The top panels show CMDs in SA~103: all SDSS data (left), Besancon model (middle) and SDSS data  for
star with proper motions (right). The gray line in the left panel shows the fiducial sequence of globular cluster M 53.
The red symbol represents RR~167, the dashed line indicates the faint limit of the proper-motion data.
The rectangular box shows the CMD-selection of stars for which proper motions are displayed in the bottom panels.
The bottom panels show the proper-motion distributions: our data (left), our data mapped with logarithmic contours 
of the proper-motion density distribution (middle), and the Besancon model distribution (right) with same contours
as for our data.}
\end{figure}

\begin{figure}
\includegraphics[scale=.55]{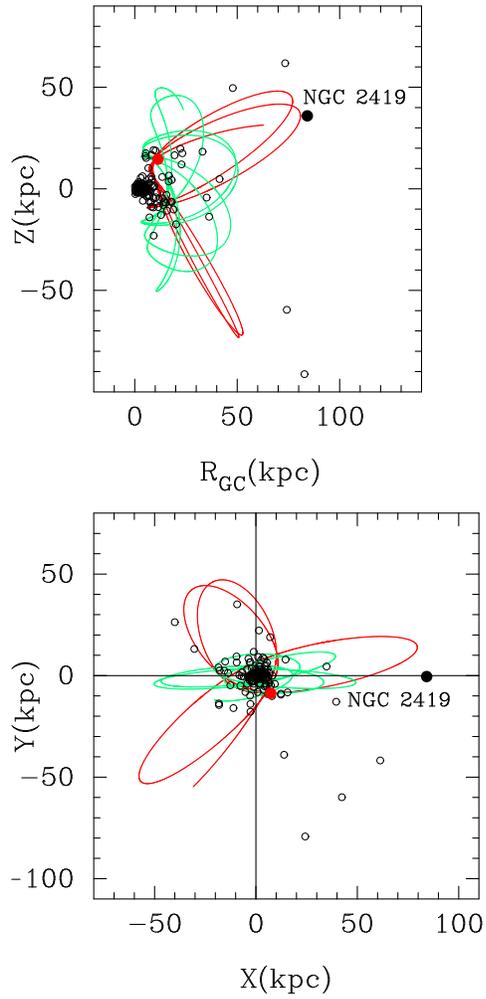}
\caption{Orbits of the VSS (red) and Sgr (green), seen in R-Z projection (left) and
in the disk plane (right) integrated back in time for 5 Gyr. Open circles 
show the globular clusters. 
NGC 2419, tentatively associated with the VSS is labeled.
The Sun is at (X,Y,Z) = (8,0,0) kpc. The current location of RR~167 is marked with a red filled circle.}
\end{figure}

\end{document}